\documentstyle[prd,aps,epsf,12pt]{revtex}
\begin{document} 

\tighten
\draft
\preprint{DAMTP96-112} 

\title{Geometry of Thermodynamic States} 
 
\author{
Dorje C. Brody$^{*}$ 
and 
Lane P. Hughston$^{\dagger}$ 
} 
\address{$*$Department of Applied Mathematics and 
Theoretical Physics,  \\ University of Cambridge,  
Silver Street, Cambridge CB3 9EW U.K.} 
\address{$\dagger$ Merrill Lynch International, 
25 Ropemaker Street, London EC2Y 9LY U.K. \\ 
and King's College London, The Strand, London 
WC2R 2LS, U.K.} 

\date{\today} 

\maketitle 

\begin{abstract} 
A novel geometric formalism for statistical estimation is applied 
here to the canonical distribution of classical statistical 
mechanics. In this scheme thermodynamic states, or equivalently, 
statistical mechanical states, can be characterised concisely in 
terms of the geometry of 
a submanifold ${\cal M}$ of the unit sphere ${\cal S}$ in 
a real Hilbert space ${\cal H}$. The measurement of a thermodynamic 
variable then corresponds to the reduction of a state vector in 
${\cal H}$ to an eigenstate, where the transition probability is 
the Boltzmann weight. We derive a set of uncertainty relations for 
conjugate thermodynamic variables in the equilibrium thermodynamic 
states. These follow as a consequence of a striking thermodynamic 
analogue of the Anandan-Aharonov relations in quantum mechanics. 
As a result we are able to provide a resolution to the controversy 
surrounding the status of `temperature fluctuations' in the 
canonical ensemble. By consideration of the curvature of the 
thermodynamic trajectory in its state space we are then able to 
derive a series of higher order variance bounds, which we calculate 
explicitly to second order. \par 
\end{abstract} 

\pacs{PACS Numbers : 05.20.Gg, 05.70.Ce, 02.40.Ky, 02.50.Cw} 


By a statistical model ${\cal M}$ we mean a family of probability 
distributions characterised by a set of parameters known as the 
parameter space. This possesses natural geometrical properties 
induced by the embedding of the family of probability distributions 
in the space of square-integrable functions on the relevant 
sample space. More precisely, by consideration of the 
parameterised square-root density function we can regard the 
space ${\cal M}$ as a submanifold of the unit sphere ${\cal S}$ 
in a real Hilbert space ${\cal H}$. Therefore, ${\cal H}$ embodies 
the state space of the system, and the properties of the statistical 
model can be described in terms of the embedding of ${\cal M}$ in 
${\cal H}$. The geometry thus arising possesses a natural 
Riemannian metric, the Fisher-Rao metric, and as a consequence 
of this the powerful tools of differential geometry can be applied 
to various aspects of statistical inference. \par 

To what extent is this methodology applicable to statistical physics? 
In the present Letter, we focus on the statistical model ${\cal M}$ 
that constitutes the state space of classical statistical mechanics, 
for which the corresponding probability density is given by the 
Gibbs distribution. By taking the square root of this density, we can 
map, for each value of the parameter, the associated probability 
density to a vector in ${\cal H}$. We are thus led to formulate a 
classical theory of measurement and statistical estimation in the 
language of Hilbert space geometry, with applications to classical 
and quantum statistical mechanics. This formulation of statistical 
theory ties up with the generalised probability theory due to Segal 
\cite{segal}. Our aim here, however, is to formulate matters in a 
geometric manner, in such a way that the statistical operations 
associated with estimation problems become more transparent. \par 

As a consequence, we find that a classical thermodynamic state, in 
the energy representation, can be expressed as a real superposition 
of energy eigenstates, each having the square-root of the 
corresponding Boltzmann weight as coefficient. The temperature 
estimation in such equilibrium states suffers from an intrinsic 
uncertainty, leading to what we might call a thermodynamic 
uncertainty relation. The source of the uncertainty can be said to 
be thermal noise, which is analogous to the quantum noise that gives 
rise to the Heisenberg uncertainty relations. Indeed, quantum 
mechanical Schr\"odinger trajectories can be obtained by Wick 
rotating the corresponding classical thermal trajectories on 
${\cal S}$, and hence we can enquire to what extent standard 
quantum mechanical relations have analogues in a classical 
thermodynamic system. We shall demonstrate, for example, that the 
classical relationship between the heat capacity of a system in 
thermal equilibrium and the energy variance in the associated Gibbs 
distribution is in direct correspondence with the Anandan-Aharonov 
relation \cite{aa} in quantum mechanics, which associates the 
velocity of a quantum mechanical state trajectory with the energy 
uncertainty along it. From there we are led to a series of 
higher order variance bounds on the temperature measurement, which 
also have quantum mechanical analogues \cite{dblh1}. \par 

Consider a real Hilbert space ${\cal H}$ with a symmetric inner 
product $g_{ab}$. A probability density function $p(x)$ can be mapped 
into ${\cal H}$ by taking the square-root $\psi(x) = \sqrt{p(x)}$, 
which is denoted by a vector $\psi^{a}$ in ${\cal H}$. The 
normalisation condition $\int (\psi(x))^{2}dx=1$, written 
$g_{ab}\psi^{a}\psi^{b}=1$, indicates that $\psi^{a}$ lies on the 
unit sphere ${\cal S}$ in ${\cal H}$. Since a probability density 
function is nonnegative, the image of the mapping 
$f: p(x)\rightarrow\psi(x)$ is the intersection 
${\cal S}_{+}={\cal S}\cap{\cal H}_{+}$ of the 
sphere ${\cal S}$ in ${\cal H}$ with the convex cone ${\cal H}_{+}$ 
formed by the totality of quadratically integrable nonnegative 
functions. We call $\psi^{a}$ the state vector of the corresponding 
probability density $p(x)$. \par 

A typical random variable is represented on ${\cal H}$ by a symmetric 
tensor $X_{ab}$, whose expectation in a normalised state 
$\psi^{a}$ is $E_{\psi}[X]=X_{ab}\psi^{a}\psi^{b}$. Similarly, the 
expectation of its square is $X_{ac}X^{c}_{b}\psi^{a}\psi^{b}$. 
The variance of $X_{ab}$ in the state $\psi^{a}$ is therefore 
${\rm Var}_{\psi}[X] = {\tilde X}_{ac}{\tilde X}^{c}_{b} 
\psi^{a}\psi^{b}$, where ${\tilde X}_{ab} = X_{ab} - g_{ab} 
E_{\psi}[X]$ represents the deviation of $X_{ab}$ from its mean 
in the state $\psi^{a}$. \par 

Suppose we have a family of probability distributions that are 
conditioned to a set of parameters $\theta$, with density function 
$p(x,\theta)$. Then, for each value of $\theta$ we obtain a 
corresponding point on ${\cal S}$ given by $\psi^{a}(\theta)$. 
This conditioning is characterised by the specification of a 
submanifold ${\cal M}$ in ${\cal S}$. 
Assuming $\psi^{a}(\theta)$ has continuous second derivatives, we 
find that ${\cal M}$ is a Riemannian manifold, with the Fisher-Rao 
metric $G_{ij} = 4g_{ab}\partial_{i}\psi^{a}\partial_{j}\psi^{b}$, 
where $\partial_{i}=\partial/\partial\theta^{i}$. This is the metric 
induced on ${\cal M}$ by the spherical geometry of ${\cal S}$.  \par 

In a statistical mechanical context, the parametrised 
family of probability distribution takes the form of 
the Gibbs measure 
\begin{equation} 
p(x,\theta)\ =\ q(x) \exp\left[- \sum_{j}\theta^{j}H_{j}(x) 
- W_{\theta} \right] \ , \label{eq:gib} 
\end{equation} 
where the variable $x$ ranges over the configuration space, 
$H_{j}(x)$ represents the form of the energy, $W_{\theta}$ is a 
normalisation factor, and $q(x)$ determines the distribution at 
$\theta^{j}=0$. Our goal is to formulate a Hilbert space 
characterisation of this distribution. In fact, it can be shown 
\cite{dblh1} that the 
state vector $\psi^{a}(\theta)$ in ${\cal H}$ corresponding to 
the Gibbs distribution (\ref{eq:gib}) satisfies the differential 
equation 
\begin{equation} 
\frac{\partial\psi^{a}}{\partial\theta^{j}}\ =\ - 
\frac{1}{2}{\tilde H}^{a}_{j b} \psi^{b}\ , \label{eq:thermo} 
\end{equation} 
where ${\tilde H}_{j ab}=H_{j ab}-g_{ab}E_{\psi}[H_{j}]$. The 
solution of this equation is 
\begin{equation} 
\psi^{a}(\theta) = \exp\left[- \frac{1}{2} \left( 
\sum_{j}\theta^{j}H^{a}_{j b} + {\tilde W}_{\theta}\delta^{a}_{b}
\right) \right] q^{b}\ , 
\end{equation} 
where ${\tilde W}_{\theta} = W_{\theta}-W_{0}$ and $q^{a} = 
\psi^{a}(0)$ is the prescribed distribution at $\theta^{j}=0$. 
The Fisher-Rao metric for the parameter 
space ${\cal M}$ of the Gibbs distribution is 
$G_{ij} = \partial_{i}\partial_{j}W_{\theta}$. Thus, for models in 
statistical mechanics where the normalisation is known, or 
equivalently, for which the partition function is specified, one 
can study the geometry of ${\cal M}$ directly by use of this 
expression \cite{stat}. \par 

We are now in a position to set up a microscopic theory of 
measurement for thermodynamic systems. For simplicity, we 
consider a one parameter family of thermal states, letting 
$\beta = 1/k_{B}T$ denote the usual inverse-temperature parameter 
and $H_{ab}$ the symmetric Hamiltonian for the system. For the 
state vector we have 
\begin{equation} 
\psi^{a}(\beta)\ =\ \exp\left[ -\frac{1}{2}(\beta H^{a}_{b} + 
{\tilde W}_{\beta}\delta^{a}_{b})\right] q^{b}\ , \label{eq:psi1} 
\end{equation} 
and for each value of the 
temperature we find a point on ${\cal M}$ in ${\cal S}$. To be 
more specific, we have a unit sphere ${\cal S}$ in ${\cal H}$, 
whose axes label the configurations of the system, each of which 
has a definite energy. Suppose we 
let $u^{a}_{k}$ denote an orthonormal basis of vectors in 
${\cal H}$. Here, the index $k$ labels all the 
points in the phase space of the given statistical system. In other 
words, for each point in phase space we have a 
corresponding basis vector $u^{a}_{k}$ in ${\cal H}$ for some 
value of $k$. With this choice of basis, a classical thermal 
state $\psi^{a}(\beta)$ can be expressed as a superposition 
\begin{equation} 
\psi^{a}(\beta)\ =\ e^{-\frac{1}{2}W_{\beta}} 
\sum_{k} e^{-\frac{1}{2}\beta E_{k}} u^{a}_{k}\ , 
\label{eq:psi} 
\end{equation} 
where $E_{k}$ is the energy for $k$-th configuration, and 
thus $\exp[W_{\beta}]=\sum_{k}\exp(-\beta E_{k})$ is 
the partition function. The index $k$ is formal here 
in the sense that the summation may if 
appropriate be replaced by an integration. By comparing equations 
(\ref{eq:psi1}) and (\ref{eq:psi}), we find that the $\beta = 0$ 
thermal state $q^{a}$ is given by 
\begin{equation} 
q^{a}\ =\ e^{-\frac{1}{2}W_{0}} \sum_{k} u^{a}_{k}\ , 
\end{equation} 
which corresponds to the centre point in ${\cal S}_{+}$. This relation 
reflects the fact that all configurations are equally likely to 
occur at infinite temperature, with probability $\exp(-W_{0})$. 
The state trajectory $\psi^{a}(\beta)$ 
thus commences at the centre point $q^{a}$, and follows a curve 
on ${\cal S}$ generated by the Hamiltonian $H_{ab}$, for which the 
curvature is 
\begin{equation} 
K_{\psi}(\beta)\ =\ \frac{\langle {\tilde H}^{4}\rangle}
{\langle{\tilde H}^{2}\rangle^{2}} 
- \frac{\langle{\tilde H}^{3}\rangle^{2}}
{\langle {\tilde H}^{2}\rangle^{3}} - 1\ , \label{eq:cur} 
\end{equation} 
where $\langle {\tilde H}^{n}\rangle$ denotes the $n$-th central 
moment of the observable $H_{ab}$. Here as usual the curvature of 
the curve $\psi^{a}(\beta)$, which is necessarily positive, 
is defined by $K_{\psi}(\beta)=g_{ab}\psi^{a}_{2} 
\psi^{b}_{2}/(g_{ab}\psi^{a}_{1}\psi^{b}_{1})^{2}$, where 
$\psi^{a}_{2} = \ddot{\psi}^{a}-\dot{\psi}^{a}{\ddot \psi}^{b}
\dot{\psi}_{b}/\dot{\psi}^{c}\dot{\psi}_{c} - \psi^{a}
\ddot{\psi}^{b}\psi_{b}$ is the `acceleration' vector along 
$\psi^{a}(\beta)$, and $\psi^{a}_{1}=\dot{\psi}^{a}$ is the 
`velocity' (the dot denotes $\partial/\partial\beta$). The 
acceleration satisfies $\psi^{a}_{2}\psi_{a}=0$ 
and $\psi^{a}_{2}\psi_{1a}=0$. Note that $\psi^{a}_{1}\psi_{a}=0$ 
since $\psi^{a}(\beta)$ lies on ${\cal S}$. \par 

If the dimension of ${\cal H}$ is infinite, the statistical 
model may exhibit a phase transition at a critical point $\beta_{c}$. 
Then the curve proliferates into $L$ distinct curves, where $L$ is 
the multiplicity of the ground state degeneracy. Thus, for the 
thermal states it is important to fix the initial condition at 
$\beta=0$, since any other point can be ambiguous. Physically, 
this ambiguity reflects the various coexisting phases allowed at the 
critical point. In particular, if the transition is of second order, 
the curvature is singular at $\beta_{c}$. It follows from the 
expression (\ref{eq:cur}) that the scaling behaviour of $K_{\psi}$ 
around $\beta_{c}$ is given by 
$K_{\psi} \sim |{\hat \beta}|^{-\kappa}$, where ${\hat \beta} = 
\beta_{c}/\beta-1$ is the reduced temperature and $\kappa = 2-\alpha$ 
in terms of conventional critical exponents. The standard relation 
$2-\alpha = d\nu$ indicates that the curvature scales like 
correlation volume \cite{stat}. \par 

The simplest model for a measurement of the state can be 
described by projecting out a point in phase space. The 
resulting probability for observing the state 
$u^{a}_{k}$ for some value of $k$ is thus given by the 
Boltzmann weight 
\begin{equation} 
p_{k}\ =\ (g_{ab}\psi^{a}u^{b}_{k})^{2}\ =\ 
e^{-\beta E_{k}-W_{\beta}}\ . 
\end{equation} 
This model can be extended to incorporate probability operator 
valued measures. For instance, suppose we are interested in the 
measurement of an observable $X_{ab}$ for a state $\psi^{a}(\beta)$. 
Then, the probability density for the measurement outcome $x$ is 
the expectation $p(x,\beta) = \Pi_{ab}(X,x)\psi^{a}\psi^{b}$ of the 
projection operator 
\[ 
\Pi^{a}_{b}(X,x)\ =\ \frac{1}{\sqrt{2\pi}} 
\int_{-\infty}^{\infty} \exp\left[ i\lambda(X^{a}_{b} - 
x\delta^{a}_{b}) \right] d\lambda  \ . 
\]  
The choice of $\Pi_{ab}$ can also 
include nonorthogonal resolution operators. \par 

Now let us turn to the measurement problem for thermal states. 
Our intention is to study the uncertainties arising in making 
inferences from the measurement outcomes. In general, we may wish 
to estimate the value of a function of an unknown parameter, such as 
internal energy or magnetic susceptibility. We shall consider, in 
particular, the case when we estimate the temperature. Suppose that 
$B_{ab}$ is an unbiased estimator for the parameter $\beta$, so 
$B_{ab}\psi^{a}\psi^{b}/g_{cd}\psi^{c}\psi^{d} = \beta$. 
Then, the variance in estimating $\beta$ can be expressed 
\cite{dblh2,dblh3} by the geometrical relation 
\begin{equation} 
{\rm Var}_{\psi}[B]\ =\ \frac{1}{4} 
g^{ab}\nabla_{a}\beta\nabla_{b}\beta\ , 
\end{equation} 
on the unit sphere ${\cal S}$, where $\nabla_{a}\beta = 
\partial\beta/\partial\psi^{a}$ is the gradient of the 
temperature estimate $\beta$. 
The essence of this relation can be understood as follows. 
First, recall that $\beta$ is the expectation of the operator 
$B_{ab}$ in the 
state $\psi^{a}(\beta)$. Suppose that the state changes rapidly 
as $\beta$ changes. Then, the variance in estimating $\beta$ is 
small, and indeed, this is given by the squared 
magnitude of the `functional derivative' of $\beta$ with respect 
to the state $\psi^{a}$. On the other hand, if the state does not 
change significantly as $\beta$ changes, then the measurement 
outcome of an observable is less conclusive in determining the 
value of $\beta$. A crude example is as follows. Suppose we infer 
the value of the temperature for a magnetic system from 
measurements of the magnetisation. If the measurement outcome is, 
say, close to zero, then the temperature can be any value above the 
Curie point, and the variance is large. \par 

The squared length of the gradient vector $\nabla_{a}\beta$ can be 
expressed as a sum of squares of orthogonal components. To this 
end, we choose a new set of orthogonal basis vectors 
given by the state $\psi^{a}$ and its higher order derivatives. 
If we let $\psi^{a}_{n}$ denote $\psi^{a}$ for $n=0$, and for 
$n>0$ the component of the derivative $\partial^{n}\psi^{a}/
\partial\beta^{n}$ orthogonal to the state $\psi^{a}$ and its 
lower order derivatives, then our orthonormal vectors are 
given by ${\hat \psi}^{a}_{n} = 
\psi^{a}_{n}(g_{bc}\psi^{b}_{n}\psi^{c}_{n})^{-1/2}$ 
for $n=0,1,2,\cdots$. 
The reason for choosing this set instead of the original basis 
$u^{a}_{k}$ is for computational simplicity. With this 
choice of orthonormal vectors, we find that the variance of the 
estimator $B$ satisfies the inequality  
\begin{equation} 
{\rm Var}_{\psi}[B]\ \geq\ \sum_{n} 
\frac{({\tilde B}_{ab}\psi^{a}_{n}\psi^{b})^{2}}
{g_{cd}\psi^{c}_{n}\psi^{d}_{n}} \ , \label{eq:vb} 
\end{equation} 
for any range of the index $n$. This follows as a consequence of 
the fact that the squared magnitude of the vector $\frac{1}{2} 
\nabla_{a}\beta = {\tilde B}_{ab} \psi^{b}$ is necessarily greater 
than or equal to the sum of the squares of its projections onto the 
basis vectors given by ${\hat \psi}^{a}_{n}$ for the specified 
range of $n$. \par 

In particular, for $n=1$ we have $B_{ab}\psi^{a}_{1}\psi^{b} = 
\frac{1}{2}$ on account of the relation $B_{ab}\psi^{a}\psi^{b}= 
\beta$, and $g_{ab}\psi^{a}_{1}\psi^{b}_{1}=\frac{1}{4} 
\Delta H^{2}$, which follows from the differential equation 
$\partial\psi^{a}/\partial\beta=-\frac{1}{2}{\tilde H}^{a}_{b}
\psi^{b}$. Therefore, as a consequence of equation (\ref{eq:vb}), 
if we write ${\rm Var}_{\psi}[B] = \Delta\beta^{2}$, we find for 
$n=1$ the following thermodynamic uncertainty relation: 
\begin{equation} 
\Delta\beta^{2}\Delta H^{2}\ \geq\ 1 \label{eq:tu} 
\end{equation} 
which is valid along the trajectory ${\cal M}$ consisting of 
the thermal equilibrium states $\psi^{a}(\beta)$. Interestingly, 
Landau and Lifshitz have obtained an inequality of this kind on 
the basis of temperature fluctuations. Kittel and Kroemer argue, 
on the other hand, that such an inequality is meaningless, since the 
temperature is a fixed constant by definition in the canonical 
distribution. The framework we have presented provides a 
mathematically and physically consistent solution to this long 
standing point of controversy \cite{llk}. That is, $\beta$ is 
indeed a fixed constant for a canonical ensemble, which does not 
fluctuate. However, for a given equilibrium system, if we wish to 
find the actual value of $\beta$, there is an inevitable uncertainty 
associated with our estimation, characterised by (\ref{eq:tu}).\par 

While the variance bounds in (\ref{eq:vb}) formally depend on the 
specific choice of the estimator $B_{ab}$, we find, remarkably, that 
in the case of thermal states, these bounds are systematically 
independent of the estimator $B$. For example, for $n=2$, the 
acceleration vector $\psi^{a}_{2}$ is given by the expression 
\[ 
\psi^{a}_{2}\ =\ \frac{1}{4}\left( 
{\tilde H}^{a}_{b}{\tilde H}^{b}_{c}\psi^{c} - \frac{\langle 
{\tilde H}^{3}\rangle}{\langle {\tilde H}^{2}\rangle}
{\tilde H}^{a}_{b}\psi^{b} - \langle{\tilde H}^{2}\rangle\psi^{a} 
\right)\ . 
\] 
To value the corresponding correction term in (\ref{eq:vb}) we 
note that ${\tilde B}_{ab}\psi^{a}_{2}\psi^{b}=B_{ab}\psi^{a}_{2}
\psi^{b}$ on account of the orthogonality $g_{ab}\psi^{a}_{2}\psi^{b} 
= 0$. Then, by differentiating the relation $B_{ab}\psi^{a}\psi^{b} 
= \beta$, we obtain $B_{ab}{\tilde H}^{b}_{c}\psi^{a}\psi^{c}=-1$. 
Taking a second derivative we conclude that $B_{ab}{\tilde H}^{a}_{c} 
{\tilde H}^{b}_{d}\psi^{c}\psi^{d} = 
\beta\langle{\tilde H}^{2}\rangle$, providing that $B_{ab}$ commutes
with $H_{ab}$, which is the case for classical thermodynamic 
variables. Substitution of these relations into (\ref{eq:vb}) yields 
a sharper thermodynamic uncertainty relation 
\begin{equation} 
\Delta\beta^{2}\Delta H^{2}\ \geq\ 1 + 
\frac{\langle{\tilde H}^{3}\rangle^{2}}{\langle{\tilde H}^{2}
\rangle^{3} K_{\psi}} \ , \label{eq:gtu} 
\end{equation} 
where the expression for the curvature $K_{\psi}$ is given 
in (\ref{eq:cur}). It is clear that, by consideration of other 
equilibrium distributions such as grand canonical or $P$-$T$ 
distributions, analogous relations can be derived, e.g., for 
chemical potential $\mu/k_{B}T$ and particle number $N$, 
or pressure $P/k_{B}T$ and volume $V$. 
The equality in (\ref{eq:tu}) holds if the energy expectation 
is proportional to the inverse temperature $\beta$. 
It is also interesting to observe that by virtue of the 
formula $2B_{ab}\psi^{a}_{1}\psi^{b}=1$, these conjugate 
variables satisfy the covariance relation $E_{\psi}[BH] - 
E_{\psi}[B]E_{\psi}[H] = - 1$, which can be expressed in the 
form of the anticommutation relation $\{ B, {\tilde H}\}_{\psi} 
= -1$, valid in expectation along the thermal trajectory. \par   

We note, incidentally, that the Fisher-Rao metric $G=4g_{ab} 
\psi^{a}_{1}\psi^{b}_{1}$ in this case is given by $\Delta H^{2} 
= T^{2}C$, where $C$ is the heat capacity. For systems exhibiting 
second order phase transitions, the heat capacity diverges at the 
critical point. Therefore, the 
temperature uncertainty $\Delta \beta^{2}$ can be made small in the 
vicinity of such critical points. This is due to the sensitivity of 
the state near critical points. Conversely, it is clear that 
temperature estimation becomes difficult for values of $\beta$ 
far from the critical value $\beta_{c}$. The relation $G = 
\Delta H^{2}$ is the thermodynamic counterpart of the Anandan-Aharonov 
relation \cite{aa} in geometric quantum mechanics. However, unlike the 
quantum case where $G$ is constant along Schr\"odinger trajectories, 
for thermal trajectories $G$ depends upon the parameter $\beta$. In 
particular, we find that $\partial_{\beta}G = 2 
\langle{\tilde H}^{3}\rangle$, which shows that for the higher order 
correction given in (\ref{eq:gtu}) to be nontrivial, it suffices that 
the bound in (\ref{eq:tu}) should not be saturated. \par 

In the foregoing sketch of our theory of statistical measurement, 
we have observed that many standard quantum mechanical 
operations are already present at an essentially classical level of 
probabilistic reasoning. 
This is surprising, since the general view in physics is that 
the Hilbert space structure associated with the space of states 
in nature is special to quantum theory, and has 
no analogue in classical probability theory and statistics. 
Therefore, we can ask to what extent quantum theory is 
distinguished from the classical probability 
theory discussed above. From the viewpoint of Hilbert space 
geometry, indeed, the quantum theory can be regarded in a certain 
sense as a special case of the 
generalised statistical theory. That is, the theory outlined 
above can be specialised to the space of quantum states 
by introducing a compatible complex structure $J^{a}_{b}$ on 
the underlying real Hilbert space ${\cal H}$. In this case, the 
state trajectory $\xi^{a}(t)$, having the time $t$ as parameter, 
satisfies the Schr\"odinger equation $\partial\xi^{a} / 
\partial t = J^{a}_{b}{\tilde H}^{b}_{c}\xi^{c}$ and substitution 
of this into equation (\ref{eq:vb}), for $n=1$, yields the quantum 
mechanical uncertainty relation $\Delta t^{2}\Delta H^{2} \geq 
1/4$. \par 

The constructions given above for the thermodynamic equilibrium 
states $\psi^{a}(\beta)$ and the Schr\"odinger trajectories 
$\xi^{a}(t)$ illustrate two examples of probabilistic theories 
that can be analysed within the framework of Hilbert space 
geometry. For nonequilibrium systems, on the other hand, 
we may need to consider statistical dynamics \cite{rs}, in which 
case the differential equation for the state $\psi^{a}$ has to be 
modified appropriately. Once a model has been chosen, the 
formalism can be applied to study the geometrical 
and statistical properties of these systems. \par 

Our approach has been to view the thermodynamic state space 
as an example that can be put forth within the framework 
of a generalised statistics, with an emphasis on the 
geometric structure of the underlying real Hilbert space. 
This generalised probability theory might at first appear to be 
formulated by way of an {\it ad-hoc} analogy with quantum theory. 
However, as we proceed, we find that the Hilbert space structure 
associated with the probability distributions is indeed the 
fundamental construction. 
In this formulation, various concepts in statistical studies, 
such as the notion of uncertainty, have precise geometric 
characterisations, hence allowing a transparent understanding 
of the underlying physics of the given model. We note that 
the higher order correction in (\ref{eq:gtu}) can easily be 
calculated for specific models. For example, for an $N$-spin 
Ising chain, we find that the correction term is given by 
$2\sinh^{2}(\beta J)/(N-1)$, where $J$ is the exchange 
integral.\par 

The authors acknowledge their gratitude to B. K. Meister and 
R. F. Streater for useful discussions. DCB is also grateful to 
PPARC for financial support. \par 

$*$ Electronic address: d.brody@damtp.cam.ac.uk \par 
$\dagger$ Electronic address: lane@ml.com\par

\end{document}